\documentclass[10pt,conference]{IEEEtran}
\usepackage{amsmath, amssymb, graphicx, cite, color}
\usepackage{algorithmic}

\newtheorem{theorem}{Theorem}
\newtheorem{lemma}{Lemma}
\newtheorem{proposition}{Proposition}
\newtheorem{corollary}{Corollary}

\newtheorem{definition}{Definition}

\newcommand{\be}{\begin{eqnarray}}
\newcommand{\ee}{\end{eqnarray}}

\newcommand{\EX}{\mathbf{E}}
\newcommand{\PR}{\mathbf{P}}

\newcommand{\mc}{\mathcal}
\newcommand{\mbf}{\mathbf}

\newcommand{\GI}{\mathsf{GI}}
\newcommand{\Mq}{\mathsf{M}}
\newcommand{\Dq}{\mathsf{D}}

\IEEEoverridecommandlockouts 

\title{Qubits through Queues: The Capacity of Channels with Waiting Time Dependent Errors}
\author{
\IEEEauthorblockN{ Avhishek Chatterjee, Krishna Jagannathan}\thanks{Authors are listed alphabetically.}
\IEEEauthorblockA{Department of Electrical Engineering\\IIT Madras, Chennai, India}\and
\IEEEauthorblockN{Prabha Mandayam}
\IEEEauthorblockA{Department of Physics\\IIT Madras, Chennai, India}
}

\begin{document}

\maketitle

\begin{abstract}
We consider a setting where qubits are processed sequentially, and derive fundamental limits on the rate at which classical information can be transmitted using quantum states that decohere in time. Specifically, we model the sequential processing of qubits using a single server queue, and derive explicit expressions for the capacity of such a `queue-channel.' We also demonstrate a sweet-spot phenomenon with respect to the arrival rate to the queue, i.e., we show that there exists a value of the arrival rate of the qubits at which the rate of information transmission (in bits/sec) through the queue-channel is maximized. Next, we consider a setting where the average rate of processing qubits is fixed, and show that the capacity of the queue-channel is maximized when the processing time is deterministic. We also discuss design implications of these results on quantum information processing systems.
\end{abstract}

%\begin{IEEEkeywords}
%channel capacity, quality of service, queuing 
%\end{IEEEkeywords}

\section{Introduction}
\label{sec:intro}
Quantum bits (or qubits) have a tendency to undergo rapid decoherence in time, due to certain fundamental physical phenomena. The manner and mechanism of such decoherence depends on the underlying physical implementation of the quantum state, the environment in which the quantum state  evolves, and other physical factors such as temperature. Once a state decoheres, the information stored is lost either partially or completely, depending again on the underlying realizations and physical processes.

In this paper, we are concerned with \emph{sequential processing of a stream of qubits} --- for example, this can include transmitting, storing or performing gate operations on the quantum states. In this setting, we derive fundamental bounds on the rate at which information can be conveyed using quantum states that decohere in time. 

When quantum states are prepared and then processed sequentially, it is reasonable to posit that there will inevitably be a non-zero `processing time,' corresponding to each qubit, which in turn corresponds to a finite rate at which the qubits can be processed by the system. For example, when the qubit is prepared and transmitted as a photon polarization state, %it loses fidelity as it traverses a certain length of fiber (or free space) before it is processed at the receiver end. 
the rate at which a receiver can detect (and hence process) the photons is constrained by the average dead-time of the detectors, which is typically of the order of a tens of nanoseconds~\cite{hadfield2009single}. To consider another concrete example, superconducting Josephson junction based qubits have gate processing times ranging from a few tens of nanoseconds to a few hundreds of nanoseconds, while their average coherence times are typically of the order of a few tens of microseconds~\cite[Table 2]{wendin2017quantum}. In such a scenario, the coherence time of each qubit is only about two to three orders of magnitude longer than the time it takes to process each qubit. This brings us to an interesting phenomenon, which does not seem to arise naturally in transmitting classical bits:  as the qubits wait to be processed, they inevitably undergo decoherence, leading to errors. 
\begin{figure}
  \centering
  \includegraphics[width=3.5in]{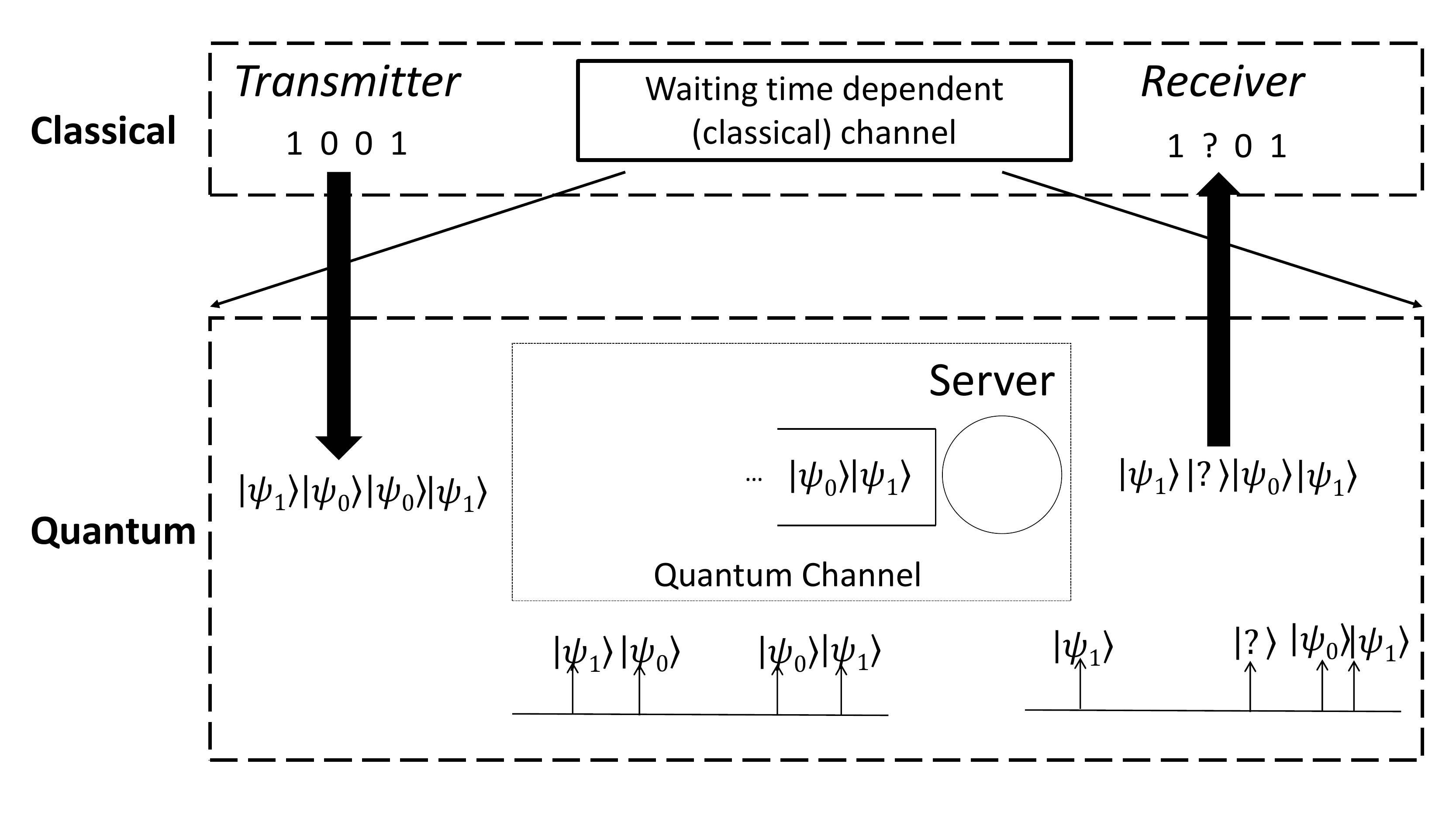}
  \caption{Schematic of the queue-channel depicting the case of quantum erasure.}
  \label{fig:diagram}
\end{figure}

More generally, we may consider a setting where qubits are prepared (or ``arrive'') according to some random process at a particular rate, and are to be processed sequentially. Due to the non-zero processing time for each qubit, the arriving qubits will have to wait in sequence to be processed. %In the aforementioned example, this may be achieved via an optical delay line. 
The present paper focusses on obtaining a quantitative characterization of the above phenomenon. Specifically, we model the sequential processing of qubits using a single-server queue with average service rate $\mu$. Now, suppose that the qubits `arrive' at the queue according to a stationary random process of rate $\lambda.$ Since the queue is stable if and only if $\lambda<\mu,$ it is immediately clear that this system cannot process qubits at a rate higher than $\mu.$ A key question we address in this paper is as follows: Assuming for simplicity that each qubit is used to encode one classical bit, is it possible to transmit information through the above queue at a rate that is arbitrarily close to $\mu$ bits/sec? 

In the case of classical bits, the answer is clearly in the affirmative. However, in the quantum case, we show that the answer turns out to be in the negative in general. Intuitively, when the arrival rate $\lambda$ is very close to $\mu,$ the waiting time for each qubit becomes very large. As a result, most of the qubits are likely to suffer decoherence, which leads to a higher probability of error.  %A functional abstraction of a system for sequential processing of qubits is presented in\ref{fig:diagram}. 

Indeed, under physically well-motivated models for the decoherence of qubits with time, we derive explicit expressions for the capacity of the above `queue-channel\footnote{A terminology we borrow from \cite{ChatterjeeSV2017}.}.' In particular, we demonstrate a `sweet-spot' phenomenon with respect to the arrival rate, i.e., we show that there  exists a particular value of arrival rate $\lambda^*\in (0,\mu)$ at which the rate of information  transmission (in bits/sec) through the queue-channel is maximized. 

Next, for a given average rate $\mu$ of processing qubits, and Poisson arrivals of qubits, we prove that the channel capacity is maximized when the processing time of each qubit is deterministic.  In other words, given a processing rate $\mu$, the rate of information transmission is maximized by ensuring that the processing time is deterministic for each qubit.

Finally, we remark that similar waiting time dependent errors can also be observed in other emerging as well as classical systems. For example, due to the short-lived nature of human attention, the performance of a human deteriorates with the waiting time \cite{Kahneman1973}. In this context, a waiting time dependent channel arises due to human impatience instead of quantum decoherence. This is particularly relevant to 
crowdsourcing. In the context of age of information \cite{CostaCE2014}, packets become useless (erased) after waiting in a queue for a certain duration -- a scenario which also falls within the scope of the model we consider. 
\subsection{Related Literature}
%ADD SOME QUANTUM REFERENCE WHICH ARE NOT MATHEMATICALLY RELATED BUT APPLICATION AREA WISE FIRST

Gallager and Telatar initiated the area of multiple access queues in \cite{TelatarG1995} which is the first published work at the intersection of queuing and information theory.
Around the same time Anantharam and Verd\'{u} considered timing channels where information is encoded in the times between consecutive information packets, and these packets are subsequently processed according to some queueing discipline
\cite{AnantharamV1996}. Due to randomness in the sojourn times of packets through servers, the encoded timing information is distorted, which the receiver must decode. %Later, a discrete-time queue was studied \cite{BedekarA1998}, and  further insight was obtained by characterizing the entropy of arrival and departure processes of a queue \cite{PrabhakarG2003}.    
%Timing channels were further investigated for suitable decoding schemes \cite{SundaresanV2000}, zero-rate reliability \cite{WagnerA2005}, connections to game-theoretic settings \cite{GilesH2002}, information leakage \cite{GongKV2011,GorantlaKKCMK2012}, and models of information overload in microblogging \cite{TavanYB2013}.
In contrast to~\cite{AnantharamV1996}, \emph{we are not concerned with information encoded in the timing between packets --- in our work, all the information is in the symbols.}
 
 To the best of our knowledge, an information theoretic notion of reliability of a queuing system with state-dependent  errors was first studied in \cite{ChatterjeeSV2017}, where the authors considered queue-length dependent errors motivated mainly by human computation and crowd-sourcing.
   
%The study of information-theoretic limits of queuing multiple-access channels was pioneered by Telatar \cite{Telatar1992}, and further explored in \cite{TelatarG1995,RajTT2004}.  This line of work is essentially concerned with the reliable transmission of bursty sources \cite{MusyT2006}, as we are here.  In multiple-access settings, however, the main constraint beyond noise is interference among users.  The present work has a single user, but performance does degrade with greater burstiness,a form of self-interference as it were.  A recent study of microbial communication also had a kind of self-interference called channel clogging \cite{MichelusiBEMM2015_arXiv}.

\subsection{Contributions}
We first study a queue-channel with waiting time induced erasures %which models the decoherence of qubits 
using a quantum erasure channel \cite{bennett97}.  In the simplest $\Mq/\Mq/1$ setting, we explicitly characterize the capacity of the erasure queue-channel, and show that there is an optimal arrival rate $\lambda_{\Mq/\Mq/1}\in(0,\mu)$ at which the capacity of the queue-channel is maximized. Next, we generalize the above result to an $\Mq/\GI/1$ setting, and show a similar behaviour. 

This result highlights an unusual interplay that exists between transmission rate and delay in the quantum case. Unlike in the classical case where we can obtain any rate that is arbitrarily close to the server rate (at the expense of delay), in the quantum case, it is desirable to operate away from the server capacity from the point of view of maximizing capacity. This is because when qubits are sent faster than the optimal rate, the effective rate of information transmission actually \emph{decreases}, due to the increased waiting time induced errors. 

While the above results characterize the optimal arrival rate of qubits for a fixed service distribution, one can also ask after the best service time distribution for fixed values of arrival and service rates. Indeed, we show that the capacity of the queue-channel is maximized when the service time distribution is deterministic. In other words, the $\Mq/\Dq/1$ queue maximizes the queue-channel capacity, among all $\Mq/\GI/1$ queues. In certain physical realizations, there could be fundamental physical constraints that translate to an optimal gate processing rate of the qubits (see for example~\cite{gatefidelity16}). Our result offers an important design insight in such a scenario --- the capacity is maximised when the gate processing time is deterministic across qubits, i.e., it is desirable to mitigate `jitter' in the processing times.

Finally, we also obtain capacity results for the class of random bijective channels, which includes the quantum binary symmetric channel as an important special case.
%(i) introducing an information-theoretic model for incorporating the notion of noisy job processing in queuing systems, (ii) defining a suitable notion of capacity and proving a coding theorem to characterize it, and (iii) studying the capacity of some common queuing system models.

%Quantum state realisations using today's technology has a coherence time of blah blah...  

%As quantum computing  takes stride towards becoming an ubiquitous reality, there is a need to .... 

\section{System Model}
\label{sec:model}

The model we study is depicted in Fig.~\ref{fig:diagram}. Specifically, a source generates a classical bit stream, which is encoded into qubits. These qubits are sent sequentially to a single server queue according to a stationary point process of rate $\lambda.$ The server works like a FIFO queue with independent and identically distributed (i.i.d.)\ service times for each qubit.  After getting processed by the server, each qubit is measured and interpreted as a classical bit. We refer to this system as a queue-channel, and characterize the classical capacity of this system (in bits/sec).

In order to capture the effect of decoherence due to the underlying quantum channel, we model the error probability as an explicit function of the waiting time $W$ in the queue. For instance, in several physical scenarios, the decoherence time of a single qubit maybe modelled as an exponential random variable. In other words, the  probability of a qubit error/erasure after waiting for a time $W$ is given by $p(W)=1-e^{-\kappa W}$, where $1/\kappa$ is a characteristic time constant of the physical system under consideration~\cite[Section 8.3]{NCBook}.

\subsection{Queuing Discipline}

We consider a continuous-time system. The service requirements are i.i.d. across qubits. The service time of the $j$th qubit is denoted $S_j$ and has a cumulative distribution $F_S.$  The average service rate of each qubit is $\mu,$ i.e., $\EX_{F_S}[S] = 1/\mu.$ In the interest of simplicity and tractability, we assume Poisson arrivals, i.e., the time between two consecutive arrivals is i.i.d.\ with an exponential distribution with parameter $\lambda.$ For stability of the 
queue, we assume $\lambda<\mu$. For ease of notation let us assume $\mu=1$. (Our results easily extend to general $\mu$).

Let $A_j$ and $D_j$ be the times when $j$th qubit arrives at the queue and departs from the queue, respectively. $W_j=D_j - A_j$ be the time that $j$th qubit spends in the queue.

\subsection{Error Model}
As the qubits wait to be served, they undergo decoherence, leading to errors at the receiver. This decoherence is modelled in general as a completely positive trace preserving map \cite{NCBook}. However, in this paper, we restrict ourselves to a rudimentary setting where we use a fixed set of orthogonal quantum states (say $|\psi_0\rangle$ and $|\psi_{1}\rangle$, corresponding to classical bits $0$ and $1$, as depicted in Fig.~\ref{fig:diagram}) to encode the classical symbols at the sender's side, and measure the qubits in some \emph{fixed} basis at the receiver's end.

In general, the $j$th symbol $X_j \in \mathcal{X}$, is encoded as one of a set of orthogonal states $\{|\psi_{X_{j}}\rangle\}$ belonging to a Hilbert space $\mathcal{H}$ of dimension $\vert \mathcal{X}\vert$. The noisy output state $|\tilde{\psi}_{j}\rangle$ is measured by the receiver in some fixed basis, and decoded as the output symbol $Y_j \in \mathcal{Y}$. This measurement induces a conditional probability distribution $\PR(Y_j|X_j, W_j)$, which we can think of as an \emph{induced} classical channel from $\mathcal{X}$ to $\mathcal{Y}$.

An $n$-length transmission over the waiting time dependent queue-channel is denoted as follows. 
Inputs are $\{X_j: 1 \le j \le n\}$, channel distribution $\prod_j \PR(Y_j|X_j, W_j)$, and outputs are $\{Y_j: 1 \le j \le n\}$.

Throughout,  $Z^k = (Z_1, Z_2, \ldots, Z_k)$ denotes a $k$-dimensional vector
and $\mbf{Z}=(Z_1, Z_2, \ldots, Z_n, \ldots)$ denotes an infinite sequence of
random variables. 
Information is measured in bits and $\log$ means logarithm to the base $2$.

\section{Capacity of Queue-channel}
\label{sec:queueCapacity}
We are interested in defining and finding the information capacity of the queue-channel, which is simply the capacity of the induced classical channel defined above. As mentioned earlier, we restrict  ourselves to using a fixed set of orthogonal states to encode the classical symbols at the sender's side, and measuring in some fixed basis at the receiver's end. Under these constraints, the classical capacity of the queue-channel maybe written in terms of the inf-information rate~\cite{VerduH1994}. 

\subsection{Definitions}
Let $M, \hat{M} \in \mathcal{M}$ be the message to be transmitted and decoded, respectively. 
\begin{definition}
An $(n, \tilde{R}, T)$ code consists of the encoding function $X^n=f(M)$ and the decoding function $\hat{M} = g(X^n, A^n, D^n)$, where the cardinality of the message set $|\mathcal{M}| = 2^{n\tilde{R}}$, and for each codeword, the expected total time for all the symbols to reach to the receiver is less than $T$. 
\end{definition}
\begin{definition}
If the decoder chooses $\hat{M}$ with average probability of error less than $\epsilon$, that code is said to be $\epsilon$-achievable.
For any $0 < \epsilon < 1$, if there exists an $\epsilon$-achievable code $(n, \tilde{R}, T)$, the rate $R = \frac{\tilde{R}}{T}$ is said to be achievable.
\end{definition}
\begin{definition}
\label{def:capacity}
The information capacity of the queue-channel is the supremum of all achievable rates for a given arrival process with distribution $F_A$ and is denoted by $C(F_A)$ bits per unit time.
\end{definition}

We assume that the transmitter knows the arrival process statistics, but not the realizations before it
does the encoding.  However, depending on the application, the receiver may or may not know the realization of the arrival and the departure time of each symbol.

% % % % % % % % % % % % % % % % % % % % % % % % % % %

\begin{proposition}
\label{prop:expression}
The capacity of the queue-channel (in bits/sec) described in Sec.~\ref{sec:model} is given by
\begin{align}
C(F_A) = \lambda \sup_{\PR(\mathbf{X})}\underline{\mathbf{I}} (\mbf{X}; \mbf{Y} | \mbf{W}),
\label{eq:cap_exp2}
\end{align}
when the receiver knows the arrival and the departure time of each symbol. On the other hand, when the receiver does not have that information, the capacity is, 
\begin{align}
C(F_A) = \lambda \sup_{\PR(\mathbf{X})}\underline{\mathbf{I}} (\mbf{X}; \mbf{Y})\mbox{.} 
\label{eq:cap_exp3}
\end{align}
Here $\underbar{I}$ is the usual notation for inf-information rate \cite{VerduH1994}.
\end{proposition}

This result is essentially a consequence of the general channel capacity expresssion in~\cite{VerduH1994}. 
The following lemma which is used in the proof of Prop.~\ref{prop:expression} would be useful later.
\begin{lemma}
\label{lem:Wstationary}
Under the assumptions in Sec.~\ref{sec:model}, $\{W_j\}$ is a Markov process and has a unique limiting 
distribution $\pi$.
\end{lemma}
\begin{IEEEproof}
Follows from the stability results for $\GI/\GI/1$.
\end{IEEEproof}
Next, we present the proof of Prop.~\ref{prop:expression}.
\begin{IEEEproof}
The case where arrival and departure times of the qubits are not known follows directly from~\cite{VerduH1994} using (limiting) stationarity and ergodicity of the arrival and the departure processes 
of a $\GI/\GI/1$ queue. Note that a string of $n$ qubits see the joint channel 
$$\PR(\mbf{Y}|\mbf{X})=\int_{W^n} \prod_{j=1}^n \PR(Y_j|X_j, W_j) d\PR(W^n)$$
and the number of qubits coming out of the queue per unit time (asymptotically) is 
$\lambda$. Combining these two facts with standard information spectrum results we get the desired capacity result.

When the arrival and the departures times of the qubits are known at the receiver, the channel 
behaves like $X^n \rightarrow (Y^n, A^n, D^n)$. In that case it follows from~\cite{VerduH1994} that 
the capacity of this channel is
$$\lambda \sup_{\PR(\mathbf{X})}\underline{\mathbf{I}} (\mbf{X}; (\mbf{Y},\mbf{A},\mbf{D})).$$
Rest follow from the facts that $W_j=D_j-A_j$ for all $j$ and as no information is encoded in timings, $\mbf{X}$ is independent of $(\mbf{A}, \mbf{D})$. We present the main steps of the derivation here for the sake of completeness.

Note that  $\underline{\mathbf{I}} (\mbf{X}; (\mbf{Y},\mbf{A},\mbf{D}))$ is the limit superior in probability of the following quantity.
\begin{align}
& \ \ \frac{1}{n} \log \frac{\PR(Y^n, A^n, D^n|X^n)}{\PR(Y^n, A^n, D^n)} \nonumber \\
& =  \frac{1}{n} \log \frac{\PR(A^n, D^n|X^n) \PR(Y^n|X^n, A^n, D^n)}{\PR(A^n, D^n)\PR(Y^n|A^n, D^n)} \nonumber \\
& = \frac{1}{n} \log \frac{\PR(Y^n|X^n, A^n, D^n)}{\PR(Y^n|A^n, D^n)}. \nonumber \\
& = \frac{1}{n} \log \frac{\PR(Y^n|X^n, W^n)}{\PR(Y^n|W^n)}
\end{align}
The step before the last is due to independence of $X^n$ and $(A^n, D^n)$. Also, as per the channel model we have $(A^n, D^n) \rightarrow W^n \rightarrow Y^n$, which leads to the final expression. 
\end{IEEEproof}

\subsection{Remarks}
Before proceeding further, we note the difference between the maximum symbol throughput (number of symbols processed per unit time) and the maximum information throughput (our notion of capacity) of the queuing system studied here. The symbol throughput is the maximum rate of arrivals for which the queue is stable and hence, increases with $\lambda$ on $[0,\mu)$. On the other hand, the expression for `information throughput' has $\lambda$ as a multiplicative factor. However, this does not mean it increases with $\lambda$. In typical queuing systems, the average waiting time is increasing in $\lambda$. For quantum channels and other systems like crowd-sourcing and multimedia communication, service errors are more likely when waiting times are larger. Thus, increasing $\lambda$ also negatively impacts the inf-information term in the capacity expression. Hence there is typically an information throughput-optimal $\lambda \in (0,\mu)$. This will be clear when we discuss some particular scenarios of interest.

The capacity expression in Prop.~\ref{prop:expression} does not provide clear insights into the behaviour
of the system under different arrival and service statistics. Such insights are crucial for optimizing the tunable  operating points, arrival or service statistics, depending on the system constraints, and to ensure efficient 
operations. Our main contribution lies in deriving single letter capacity expressions  for some commonly encountered channel models.
First, we consider queue-channels with waiting induced erasures and then a class of channels which include binary symmetric channel with waiting induced errors. In both cases our goal is to understand the effects of the service process and the arrival rate on the capacity.

\section{Erasure Channels}
\label{sec:EC}

Erasure channels are ubiquitous in classical as well as quantum information theory. We consider a quantum erasure channel~\cite{bennett97} which acts on the $j$th state $|\psi_{X_{j}}\rangle$ a follows: $|\psi_{X_{j}}\rangle $ remains unaffected with probability $1 - p(W_{j})$, and is erased to a state $|?\rangle$ with probability $p(W_{j})$, where $p:[0,\infty) \to [0,1]$ is typically increasing. Such a model also
captures the communication scenarios where information packets become useless (erased) after a deadline. For such an erasure channel, a single letter expression for capacity can be obtained.

\begin{theorem}
\label{thm:queueCapacityErasure}
For the erasure queue-channel defined above, the capacity is $\lambda~\log |\mathcal{X}|~\EX_{\pi}\left[1-p(W)\right]$ bits/sec, irrespective of the receiver's knowledge of the arrival and the departure times of symbols.
\end{theorem}
\begin{IEEEproof}[Proof]
Proof uses  an upper-bound on $\underbar{I}(\mbf{X};\mbf{Y})$ in terms of unconditional sup-entropy rate and conditional inf-entropy rate and shows that
the capacity expression is an upper-bound. On the other hand, using a similar lower-bound on $\underbar{I}(\mbf{X};\mbf{Y})$ it is shown that for a choice of distribution of $\{X_n\}$ (i.i.d. uniform)  $\underbar{I}(\mbf{X};\mbf{Y})$ is more than the capacity expression. The fact that in the case of erasure channels the received symbol is either correct or erased
(never wrong) makes the knowledge of the arrival and departure times 
irrelevant. This along with ergodicity of the queue is used in reducing $n$-symbol bounds for $\underbar{I}$ to  single-letter bounds.

From properties of limit superior and inferior, 
	\begin{align*}
		\underline{\mathbf{I}} (\mbf{X}; \mbf{Y} | \mbf{W}) &\le \overline{\mbf{H}}(\mbf{Y}|\mbf{W}) - \overline{\mbf{H}}(\mbf{Y}|\mbf{X},\mbf{W}).
	\end{align*}

	Note that $\overline{\mbf{H}}(\mbf{Y}|\mbf{X},\mbf{W})$ is the lim-sup in probability of $\frac{1}{n} \log \frac{1}{\PR(Y^n|X^n,W^n)}$, i.e., the smallest $\beta \in \mathbb{R} \cup \{\pm \infty\}$ such that
	\begin{align*}
		\lim_{n \to \infty} \Pr\left[ \frac{1}{n} \log \frac{1}{\PR(Y^n|X^n,W^n)} \ge \beta + \epsilon \right] = 0
	\end{align*}
	for any $\epsilon > 0$.
	
	By the channel model, given $W_i$ and $X_i$, $Y_i$ is independent of any other variable. So, we have
	\begin{align}
		& \frac{1}{n} \log \frac{1}{\PR(Y^n|X^n,W^n)} \nonumber \\
		& = \frac{1}{n} \sum_{i=1}^n \log\frac{1}{\PR(Y_i|X_i,W_i)} \nonumber \\
		& = \frac{1}{n}\left[\sum_{i \in N^e}\log\frac{1}{\PR(Y_i|X_i,W_i)} + \sum_{i \in [n]\setminus N^e}\log\frac{1}{\PR(Y_i|X_i,W_i)}\right]\nonumber\\
		& = - \frac{1}{n} \sum_{i=1}^n \left[\mbf{1}(Y_i=\mc{E}) \log(1-p(W_i)) + \mbf{1}(Y_i\neq \mc{E}) \log(p(W_i))\right] \label{eq:erasure1}\mbox{,}
	\end{align}
	where $\mc{E}$ represents erasure.
	  \eqref{eq:erasure1} is due to the fact that $\PR(Y_i|X_i, W_i)=p(W_i)$ if $Y_i$ is an erasure, else, it is $1-p(W_i)$. By Lemma \ref{lem:Wstationary} the limit of the expression in \eqref{eq:erasure1} exists almost surely which is also the $\overline{\mbf{H}}(\mbf{Y}|\mbf{X},\mbf{W})$. Note that this is independent of the $\{X_i\}$ distribution. 
	  
	  Let us now consider upper-bounding $\overline{\mbf{H}}(\mbf{Y}|\mbf{W})$. Let $N^{\mc{E}}$ be the set of indices for which $Y_i=\mc{E}$.
	
	\begin{align}
		& \frac{1}{n} \log \frac{1}{\PR(Y^n|W^n)} \nonumber \\
		& = \frac{1}{n} \log \frac{1}{\PR(\{Y_i: i \in N^{\mc{E}}\}, \{Y_i: i \not\in N^{\mc{E}}\} |W^n)}
		 \nonumber \\
		 & = \frac{1}{n} \log \left(\frac{1}{\PR(\{Y_i: i \in N^{\mc{E}}\}|\{W_i: i \in N^{\mc{E}}\}} \right.
		 \nonumber  \\
		 &\ \ \left. \frac{1}{\PR(\{Y_i: i \not\in N^{\mc{E}}\}|\{W_i: i \not\in N^{\mc{E}}\})}\right)
		 \label{eq:erasure2} \\
		 & = \frac{1}{n}\left(- \sum_{i \in N^{\mc{E}}} \log \PR(\mc{E}|W_i) - \log \PR(\{Y_i: i \not\in N^{\mc{E}}\}|\{W_i: i \not\in N^{\mc{E}}\}) \right) \label{eq:erasure3} \\
		 & = \frac{1}{n}\left. \big(- \sum_{i \in N^{\mc{E}}} \log \PR(\mc{E}|W_i) - \right. \nonumber \\
		 & \ \left.  \log \PR(\{Y_i\neq \mc{E}:i \not\in N^{\mc{E}}\},\{Y_i: i \not\in N^{\mc{E}}\}|\{W_i: i \not\in N^{\mc{E}}\})\right. \big) \label{eq:erasure4} \\
		 & = \frac{1}{n}\left. \big(- \sum_{i \in N^{\mc{E}}} \log p(W_i) - \log \PR(\{Y_i\neq \mc{E}:i \not\in N^{\mc{E}}\}|\{W_i: i \not\in N^{\mc{E}}\}) \right. \nonumber \\
		 & \left. - \log \PR(\{Y_i: i \not\in N^{\mc{E}}\}|\{Y_i\neq \mc{E}:i \not\in N^{\mc{E}}\},\{W_i: i \not\in N^{\mc{E}}\})\right. \big) \nonumber \\
		 & = \frac{1}{n}\left. \big(- \sum_{i \in N^{\mc{E}}} \log p(W_i) - \sum_{i \not\in N^{\mc{E}}} 
		 \log (1-p(W_i)) \right. \nonumber \\
		 & \left. - \log \PR(\{Y_i: i \not\in N^{\mc{E}}\}|\{Y_i\neq \mc{E}:i \not\in N^{\mc{E}}\},\{W_i: i \not\in N^{\mc{E}}\})\right. \big) \nonumber \\
		 & = - \frac{1}{n} \big(\sum_{i=1}^n \left[\mbf{1}(Y_i=\mc{E}) \log(1-p(W_i)) + \mbf{1}(Y_i\neq \mc{E}) \log(p(W_i))\right] \nonumber \\
		 & \left. + \log \PR(\{Y_i: i \not\in N^{\mc{E}}\}|\{Y_i\neq \mc{E}:i \not\in N^{\mc{E}}\},\{W_i: i \not\in N^{\mc{E}}\})\right. \big)\mbox{.} \label{eq:erasure5}
	\end{align}
	\eqref{eq:erasure2} and \eqref{eq:erasure3} follow because, given $W_i$, the probability of a symbol getting erased is independent of anything else (including the input symbols). \eqref{eq:erasure4} follows
	because the event  $\{Y_i\neq \mc{E}:i \not\in N^{\mc{E}}\}$ includes $\{Y_i: i \not\in N^{\mc{E}}\}$.
	
As discussed before $$\frac{1}{n} \sum_{i=1}^n \left[\mbf{1}(Y_i=\mc{E}) \log(1-p(W_i))+ \mbf{1}(Y_i\neq \mc{E}) \log(p(W_i))\right]$$ has an almost sure limit (independent of $\{X_i\}$) and is equal 
to $\overline{\mbf{H}}(\mbf{Y}|\mbf{X},\mbf{W})$. 

Note that for an erasure channel, if $Y_i$ is not an erasure, $Y_i$ has the same value as that of 
$X_i$. So, for 
any joint distribution $\PR_{\mbf{X}}$ of input symbols:

\begin{align}
& \ \frac{1}{n}\left. \big(- \log \PR(\{Y_i: i \not\in N^{\mc{E}}\}|\{Y_i\neq \mc{E}:i \not\in N^{\mc{E}}\},\{W_i: i \not\in N^{\mc{E}}\})\right. \big)  \nonumber \\
& = - \frac{1}{n} \log \PR_{\mbf{X}}(\{Y_i: i \not\in N^{\mc{E}}\}|\{W_i: i \not\in N^{\mc{E}}\}) \nonumber \\
& = - \frac{n-|N^{\mc{E}}|}{n} \frac{1}{n-|N^{\mc{E}}|} \log \PR_{\mbf{X}}(\{Y_i: i \not\in N^{\mc{E}}\})\nonumber
\end{align}

Note that in the limit, by Lemma \ref{lem:Wstationary} $\frac{|N^{\mc{E}}|}{n}$ converges almost surely to $\EX[p(W)] < 1$. So, almost surely $n-|N^{\mc{E}}| \to \infty$. So as $n \to \infty$,
$\frac{1}{n-|N^{\mc{E}}|} \log \PR_{\mbf{X}}(\{Y_i: i \not\in N^{\mc{E}}\})$ can be seen as
$\frac{1}{N} \log \PR(X^N)$ for some large $N$. $p-\lim\sup$ of this quantity is upper-bounded by
$\log|\mc{X}|$. Thus we get an upper-bound of $(1-\EX[p(W)])\log|\mc{X}|$ on
$\underline{I}$.

	On the other hand, we also have
	\begin{align}
		\underline{\mathbf{I}} (\mbf{X}; \mbf{Y} | \mbf{W}) &\ge \underline{\mbf{H}}(\mbf{Y}|\mbf{W}) - \overline{\mbf{H}}(\mbf{Y}|\mbf{X},\mbf{W}). \label{eq:erasure6}
	\end{align}
	We have already derived an expression for the second term above. Let us choose the input distribution to be uniform and i.i.d. Then, from \eqref{eq:erasure5} it follows that after the cancellation of the first and the second term in \eqref{eq:erasure6} we obtain 
	$$ - \frac{n-|N^{\mc{E}}|}{n} \frac{1}{n-|N^{\mc{E}}|} \log \PR_{\mbf{X}}(\{Y_i: i \not\in N^{\mc{E}}\}).$$
	
	For uniform and i.i.d $\{X_i\}$ this expression almost surely converges to $(1-\EX[p(W)])\log|\mc{X}|$.

Thus, we derived an $\PR_{\mbf{X}}$ indepenent upper-bound on $\underline{I}(\mbf{X};\mbf{Y}|\mbf{W})$ which was matched by a partciular choice of $\PR_{\mbf{X}}$. Thus, for the erasure channel 
$$\sup_{\PR_X} \underline{I}(\mbf{X};\mbf{Y}|\mbf{W})=(1-\EX[p(W)])\log|\mc{X}|.$$

By multiplying with $\lambda$ we obtain the capacity of this channel. This completes the proof.
\end{IEEEproof}

This single letter capacity expression allows us to mine deeper insights on system design. It is well known in queuing that waiting time increases with increasing arrival rate. As $p(\cdot)$ is increasing,  so is $\EX_{\pi}\left[p(W)\right]$ in $\lambda.$ Therefore, it is apparent from the single letter expression (in Theorem~\ref{thm:queueCapacityErasure}) that capacity may not be monotonic in $\lambda$. This raises an interesting question: is there an optimal $\lambda$ at which the capacity is maximized? The answer to this question depends on the queuing dynamics. Therefore, we first attempt to understand it for the most fundamental queuing system in communication networks, the $\Mq/\Mq/1$ queue. %Note that as the exponential distribution maximizes Shannon entropy among all non-negative random variables with the same mean, in some sense, $\Mq/\Mq/1$ has most random (in Shannon entropy sense) arrival and service distributions. 
Interestingly, for the $\Mq/\Mq/1$ queue, there exists a simple characterization of the  capacity and the corresponding optimal arrival rate.
\begin{theorem}
\label{thm:bestArrMM1GenPsi}
The arrival rate that maximizes the information capacity of the $\Mq/\Mq/1$ queue-channel
is given by
 \begin{align}
%& \arg\max_{\lambda \in (0,1)} \lambda \left(1 - \frac{1-\lambda}{\lambda}\tilde{p}\left(\frac{1-\lambda}{\lambda}\right)\right) \nonumber \\ 
%& \arg\max_{\lambda \in (0,1)} \left(\lambda  - (1-\lambda)\tilde{p}\left(\frac{1-\lambda}{\lambda}\right)\right) \nonumber \\
%& 1 - \arg\max_{u \in (0,1)} \left(1 - u - u \tilde{p}\left(\frac{u}{1-u}\right)\right)\nonumber \\
& 1 - \arg\min_{u \in (0,1)} u~\left(1 + \tilde{p}\left(\frac{u}{1-u}\right)\right)
\label{eq:bestArrMM1GenPsi} 
 \end{align}
 where for any $u>0$, $\tilde{p}(u):=\int \exp(-u x) p(x) dx$ is the Laplace transform of $p(\cdot)$.
\end{theorem}

\begin{IEEEproof}[Proof]
This proof uses the exponential waiting time distribution of $\Mq/\Mq/1$ queue to relate
the capacity to Laplace transform of $p(\cdot)$. 

It is known that the waiting time in $\Mq/\Mq/1$ is distributed as $\exp\left(\frac{1-\lambda}{\lambda}\right)$ for $\mu=1$. Thus, 
\begin{align}
& \ \EX[p(W)] \nonumber \\
& = \int_{0}^{\infty} p(w) \frac{1-\lambda}{\lambda} \exp\left(\frac{1-\lambda}{\lambda}w\right)
dw \nonumber \\
& = \frac{1-\lambda}{\lambda}\tilde{p}\left(\frac{1-\lambda}{\lambda}\right). \nonumber
\end{align}

Thus, the capacity is given by
$$\lambda \left(1 - \frac{1-\lambda}{\lambda}\tilde{p}\left(\frac{1-\lambda}{\lambda}\right)\right).$$

So, the capacity maximizing arrival rate is the one that maximizes this expression.
\begin{align}
& \ \ \arg\max_{\lambda \in (0,1)} \lambda \left(1 - \frac{1-\lambda}{\lambda}\tilde{p}\left(\frac{1-\lambda}{\lambda}\right)\right) \nonumber \\ 
& \iff \arg\max_{\lambda \in (0,1)} \left(\lambda  - (1-\lambda)\tilde{p}\left(\frac{1-\lambda}{\lambda}\right)\right) \nonumber \\
& \iff 1 - \arg\max_{u \in (0,1)} \left(1 - u - u \tilde{p}\left(\frac{u}{1-u}\right)\right)\nonumber \\
& \iff 1 - \arg\min_{u \in (0,1)} u~\left(1 + \tilde{p}\left(\frac{u}{1-u}\right)\right)
\nonumber
 \end{align}
\end{IEEEproof}

In the case of quantum erasure channels \cite{bennett97}, decoherence of qubits over time gives rise to an interesting form for $p(\cdot)$, namely, $p(W)=1-\exp(-\kappa W)$, where $\kappa$ is a physical parameter. A detailed quantum physical discussion on this can be found in \cite{grassl97}. A relation of this kind between waiting time and erasure is also relevant in multimedia communication with deadlines and in the context of age of information. In these scenarios, when deadlines or maximum tolerable age of information packets are unknown, the exponential distribution (being the most entropic) serves as a reasonably good stochastic model. Such a model is captured by the above form of $p(\cdot)$. Hence, for this particular form of $p(\cdot)$, it is important to understand the capacity behaviour explicitly. 

\begin{corollary}
\label{cor:bestArrMM1}
For an erasure queue-channel with $p(W)=1-\exp(-\kappa W)$, $F_A(x)=1-\exp(-\lambda~x)$ and $F_S(x)=1-\exp(-x)$,
\begin{itemize}
\item[(i)] the capacity is given by $\frac{\lambda(1-\lambda)}{1-\alpha\lambda}$ bits/sec.
\item[(ii)] the capacity is maximized at $$\lambda_{\Mq/\Mq/1} = \frac{1}{\alpha} \left(1-\sqrt{1-\alpha}\right)=\frac{1}{1+\sqrt{1-\alpha}},$$
\end{itemize}
  where 
$\alpha=\frac{1}{1+\kappa}$.
\end{corollary}

\begin{figure}
\centering
  \includegraphics[scale=0.6]{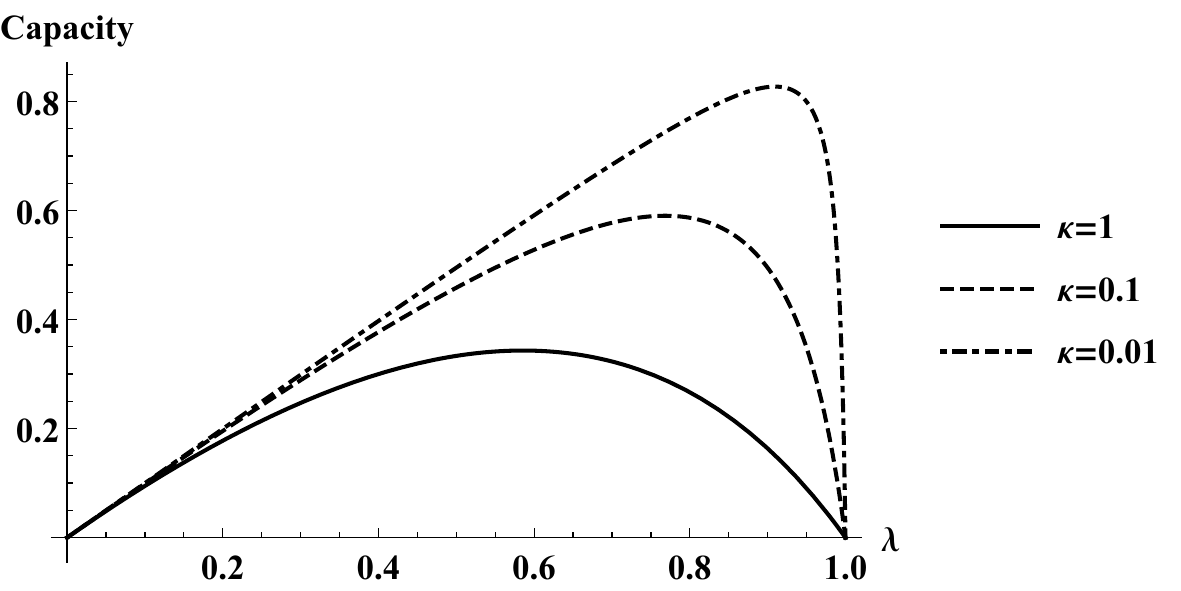}
  \caption{The capacity of the $\Mq/\Mq/1$ queue-channel (in bits/sec) plotted as a function of the arrival rate $\lambda$ for different values of the decoherence parameter $\kappa.$}
  \label{fig:plot}
\end{figure}

This result offers interesting insights into the relation between the information capacity and the characteristic time-constant of the quantum medium. In the case of decohering channels, a larger value of the decoherence exponent $\kappa$ corresponds to a faster decoherence.  We note that $\alpha$ decreases as $\kappa$ increases and hence, $\lambda_{\Mq/\Mq/1}$ decreases as $\kappa$ increases. This implies that when the qubits decohere more rapidly, the arrival rate that maximizes the capacity is lower. In other words, when the coherence time is small, it is better to send at a slower rate to avoid excessive waiting time induced errors. 

Fig.~\ref{fig:plot} depicts a capacity plot of the $\Mq/\Mq/1$ queue-channel (in bits/sec), as a function of the arrival rate $\lambda$ for different values of the decoherence parameter $\kappa.$ Since the service rate $\mu$ is taken to be unity, we note that a value of $\kappa=0.01$ corresponds to an average coherence time which is two orders of magnitude longer than the service time --- a setting  reminiscent of superconducting qubits~\cite{wendin2017quantum}. We also notice from the shape of the capacity curve for $\kappa=0.01$ that there is a drastic drop in the capacity, if the system is operated beyond the optimal arrival rate $\lambda_{\Mq/\Mq/1}.$ This is due to the drastic increase in delay induced decoherence as the arrival rate of qubits approaches the server capacity.

Next, we discuss the generalization of this result to $\Mq/\GI/1$ queues. Specifically, a result similar to Corollary~\ref{cor:bestArrMM1} also holds for $\Mq/\GI/1$ system for a different $\alpha$, though unlike for the $\Mq/\Mq/1$ queue, the waiting time is not exponentially distributed.

\begin{theorem}
\label{thm:bestArrMG1}
For an erasure queue-channel with $p(W)=1-\exp(-\kappa W)$, $F_A(x)=1-\exp(-\lambda~x)$ and a
general $F_S$ with $F_S(0)=0$, information capacity is maximized at $\lambda_{\Mq/\GI/1} = \frac{1}{\alpha} \left(1-\sqrt{1-\alpha}\right)$ for  
$\alpha=\frac{1-\tilde{F}_S(\kappa)}{\kappa}$, where $\tilde{F}_S(u)=\int \exp(-u x) dF_S(x)$.
\end{theorem}
\begin{IEEEproof}[Proof outline]
We use the Pollaczek-Khinchin formula for $\Mq/\GI/1$ queue and the fact that
the capacity is $\lambda$ times the Laplace tranform of waiting time distribution to show that the optimization of capacity over $\lambda$ is a convex problem whose optimum can be evaluated uniquely in a closed form. 

First, note that for the particular form of $p(\cdot)$, capacity is given by
$$\lambda \EX[\exp(-\kappa W)].$$

By the  Pollaczek-Khinchin formula for $\Mq/\GI/1$ queue (for $\mu=1$)
\begin{align}
& \ \EX[\exp(-\kappa W)]  \nonumber \\
& = \frac{(1-\lambda)\kappa}{\kappa - \lambda (1-\tilde{F}_S(\kappa))} \nonumber \\
& = \frac{1-\lambda}{1 - \alpha \lambda},
\end{align}
where $\alpha = \frac{(1-\tilde{F}_S(\kappa))}{\kappa}$. So, the capacity maximizing arrival rate
is
\begin{align}
\arg\max_{\lambda \in [0,1)} \frac{\lambda (1-\lambda)}{1 - \alpha \lambda}\mbox{.}\label{eq:MG1opti}
\end{align}

The following lemma (whose proof is elementary) helps us in explicitly solving (\ref{eq:MG1opti}).
\begin{lemma}
\label{lem:MG1opti}
\eqref{eq:MG1opti} is a convex optimization problem.
\end{lemma}
This lemma implies that to find the capacity maximizing $\lambda$ it is sufficient to find the $\lambda$ at
which the derivative of the capacity is $0$. Taking derivative we obtain a qudratic function in $\lambda$ which when equated to $0$ yields two solutions for $\lambda$:
$$\frac{1}{\alpha} \pm \frac{\sqrt{1-\alpha}}{\alpha}.$$
The only valid solution for which $\lambda\in[0,1)$ is given by $\frac{1}{\alpha} - \frac{\sqrt{1-\alpha}}{\alpha}$.
\end{IEEEproof}

The above results characterize an optimal $\lambda$ for given arrival and service distributions. One can also
ask after the best service distribution for a given arrival process and a fixed server rate. This question is of interest
in designing the server characteristics like gate operations \cite{gatefidelity16} or photon detectors in the case of quantum systems, and the scheduling policy in the case of packet communication with age of information constraints. The following theorem is useful in such scenarios.

\begin{theorem}
\label{thm:bestServMG1}
For an erasure queue-channel with $p(W)=1-\exp(-\kappa W)$ and $F_A(x)=1-\exp(-\lambda~x)$  at any $\lambda$ the capacity is maximized  by $F_S(x)=\mathbf{1}(x\ge 1)$, i.e., a deterministic service time maximizes capacity, among all
service distributions with unit mean and $F_S(0)=0$.
\end{theorem}
\begin{IEEEproof}[Proof]
This proof uses the relation betweeen the P-K formula for $\Mq/\GI/1$ queues and the queue-channel capacity to optimize the capacity over all service distribution
for a given $\lambda$ using Jensen's inequality. 

As derived in the proof of Theorem \ref{thm:bestArrMG1}, the capacity is
\begin{align}
& \frac{\lambda (1-\lambda)\kappa}{\kappa - \lambda (1-\tilde{F}_S(\kappa))} \nonumber \\
& = \frac{\frac{(1-\lambda)\kappa}{\lambda}}{\frac{\kappa-\lambda}{\lambda} + \tilde{F}_S(\kappa)}\mbox{.}
\nonumber
\end{align}
Thus, for any given $\lambda$, among all service distribution with unit mean, the capacity is maximized 
by that service distribution for which $\tilde{F}_S(\kappa)$ is maximized.
But by Jensen's inequality, for any service random variable $S$
$$\tilde{F}_S(\kappa)=\EX[\exp(-\kappa S)]\ge \exp(-\kappa\EX[S]).$$

So, $\tilde{F}_S(\kappa)$ is minimized by $S=\EX[S]$, i.e., a derministic service time.
\end{IEEEproof}

\section{Random Bijective and Binary Symmetric Queue-channels}

In this section, we introduce the class of random bijective queue-channels, and discuss how a single letter capacity expression can be obtained. We also discuss the binary symmetric queue-channel as a special case.

Let $\{N_i\}$ be a random sequence with values from a finite
set $\tilde{\mathcal{X}},$ and independent of the sequence
$\{X_i\}$. Also, conditioned on the sequence $\{W_i\},$ let $\{N_i\}$ be an independent (but
not identically distributed) sequence. 
From the basics of Markov dynamics, it can be shown that~\cite{MullerS2002} for an appropriate map $g:\mathcal{X}\times\tilde{\mathcal{X}} \to \mathcal{Y}$ and an
appropriate distribution of $\{N_i\}$, $\{Y_i, W_i, X_i\}$ have the same
joint distribution as $\{g(X_i,N_i), W_i, X_i\}$. Thus, instead of
$\{\PR(Y_i|W_i, X_i)\}$ the queue-channel can also be described by $\left(\tilde{\mathcal{X}}, g,\{\PR(N_i|W_i)\}\right)$.
In this section, we consider a class of queue-channels for which $\mathcal{X}=\mathcal{X}'$ and $g(X,\cdot)$ is a bijection for any $X \in \mathcal{X}$. We call such queue-channels \emph{random bijective} queue-channels. Many important class of channels, like binary (and q-ary) symmetric channels and additive noise channels belong to this class. Erasure channels are not in this class, and hence, were considered separately before. We use $N(W)$ to denote the noise random variable for a given $W$.

\begin{theorem}
\label{thm:queueCapacityRCBound}
For a random bijective queue-channel the capacity is $\lambda\left(\log |\mathcal{X}| - \EX_{\pi}\left[H(N(W))\right]\right)$ when the receiver knows  the arrival and departure times of the symbols. When the receiver does not know the arrival and departure times, the capacity is lower and upper-bounded by 
$\lambda\left(\log |\mathcal{X}| - \EX_{\pi}\left[H(\EX_{\PR(W'|W)}\left[N(W')\right])\right]\right)$ and
$\lambda\left(\log |\mathcal{X}| - H\left(\EX_{\pi}\left[N(W)\right]\right)\right)$, respectively, where
$\PR(W'|W)$ is the Markov kernel of the waiting time process $\{W_i\}$.
\end{theorem}

\begin{IEEEproof}
This result is obtained by lower- and upper-bounding inf-information rate as in the proof
of Thm.~\ref{thm:queueCapacityErasure}. But, unlike erasure channels here erroneous
symbols are received by receivers. We use the structure in the channel transitions of the reversible channels towards deriving the single letter expression.

First we consider the case without the arrival and departure times at the receiver.
Using the properties of limit superior and inferior, 
	\begin{align*}
		\underline{\mathbf{I}} (\mbf{X}; \mbf{Y}) &\le \overline{\mbf{H}}(\mbf{Y}) - \overline{\mbf{H}}(\mbf{Y}|\mbf{X}).
	\end{align*}
	Since $\overline{\mbf{H}}(\mbf{Y}) \le \log |\mathcal{X}|$ by Thm.~1.7.2 in \cite{Han2003} for any $\PR(\mbf{Y})$,
	\begin{align*}
		\underline{\mathbf{I}} (\mbf{X}; \mbf{Y}) &\le \log |\mathcal{X}| - \overline{\mbf{H}}(\mbf{Y}|\mbf{X}).
	\end{align*}
	Note that $\overline{\mbf{H}}(\mbf{Y}|\mbf{X})$ is the lim-sup in probability of $\frac{1}{n} \log \frac{1}{\PR(Y^n|X^n)}$, i.e., the smallest $\beta \in \mathbb{R} \cup \{\pm \infty\}$ such that
	\begin{align*}
		\lim_{n \to \infty} \Pr\left[ \frac{1}{n} \log \frac{1}{\PR(Y^n|X^n)} \ge \beta + \epsilon \right] = 0
	\end{align*}
	for any $\epsilon > 0$. Now
	\begin{align*}
		\frac{1}{n} \log \frac{1}{\PR(Y^n|X^n)} &= \frac{1}{n} \log \frac{1}{\PR(N_i=g^{-1}(X_i,Y_i)\ \forall i)}\mbox{,}
	\end{align*}
	since, for a random bijective channel, such a $g^{-1}$ exists for each $X_i$.
	
	As the $\{W_i\}$ process is ergodic and $\{N_i\}$ are independent given $\{W_i\}$, the liminf in probability of$ \frac{1}{n} \log \frac{1}{\PR(N^n)}$
	exists, and is equal to the entropy rate of the noise process $\{N_n\}:$
	$$H_{\infty}(\mbf{N})=\lim_{n\to\infty}H(N_{n+1}|N^n)\mbox{.}$$
	
	Therefore we obtain the converse bound that
	\begin{align*}
		\underline{\mathbf{I}} (\mbf{X}; \mbf{Y}) \le \log |\mathcal{X}| - H_{\infty}(\mbf{N})\ \ .
	\end{align*}
	
	On the other hand, we also have
	\begin{align*}
		\underline{\mathbf{I}} (\mbf{X}; \mbf{Y}) &\ge \underline{\mbf{H}}(\mbf{Y}) - \overline{\mbf{H}}(\mbf{Y}|\mbf{X}).
	\end{align*}
	The second term is $H_{\infty}(\mbf{N})$. We pick $X^n$ i.i.d.\ uniformly at random from $\mathcal{X}$. Consider $\PR(Y^n=y^n)$.
	\begin{align}
	& \PR(y^n) = \sum_{x^n} \PR(y^n,x^n) \nonumber \\
	& = \frac{1}{|\mathcal{X}|^n} \sum_{x^n} \PR(y^n|x^n) \nonumber \\
	& = \frac{1}{|\mathcal{X}|^n} \sum_{x^n}\PR(N^n=\{g^{-1}(x_i,y_i)\}) \nonumber \\
	& = \frac{1}{|\mathcal{X}|^n} \sum_{N^n}\PR(N^n) \nonumber  \\
	& = \frac{1}{|\mathcal{X}|^n}
	\end{align}
	where the step before the last follows because $g(x,\cdot)$ is bijective and hence, for a given $y$,
	$g^{-1}(\mathcal{X},y)=\mathcal{X}$. 	Thus, it follows that $\underline{\mbf{H}}(\mbf{Y})=\log |\mathcal{X}|.$ 
	
	Now let us derive upper and lower bounds for $H_{\infty}(\mbf{N})$. We note that for a given limiting marginal distribution for a process, the i.i.d. process maximizes entropy rate among all processes. This explains the upper-bound. Next, for the lower-bound, note that
	\begin{align}
	& \ H(N_{n+1}|N^n) \nonumber \\
	& \ge H(N_{n+1}|N^n, W^n) \nonumber \\
	& = - \sum_{N_{n+1}} \PR(N_{n+1},N^n, W^n) \log \PR(N_{n+1}|N^n, W^n) \nonumber \\
	& = - \sum_{N_{n+1}, W_{n+1}} \PR(N_{n+1},N^n, W^n) \nonumber \\
	 & \ \ \times \log \sum_{W_{n+1}}\PR(N_{n+1},W_{n+1}|N^n, W^n) \nonumber \\
	& = - \sum_{N_{n+1}, W_{n+1}} \PR(N_{n+1},N^n, W^n) \nonumber \\
	 & \ \ \times  \log \sum_{W_{n+1}} \PR(N_{n+1}|W_{n+1}) \PR(W_{n+1}|W_n) \nonumber \\
	& = \EX_{\PR(W_n)}\left[H(\EX_{\PR(W_{n+1}|W_n)}\left[N(W_{n+1})\right])\right] \nonumber 
	\end{align}
	
	As $\PR(W_n) \to \pi$, the lower-bound follows (since the entropy is a bounded function). Multiplying by the arrival rate $\lambda$ completes the proof.
     
     For the  case when the arrival and departure times are known, the proof is similar. Specifically, we can check that the upper-bound on $\overline{\mbf{H}}(\mbf{Y}|\mbf{W})$ 
     is $\log |\mathcal{X}|$. Similarly, for $\underline{\mbf{H}}(\mbf{Y}|\mbf{W}),$ the matching lower bound follows by choosing i.i.d. uniform distribution on $X$. 
     
     On the other hand, we have $\overline{\mbf{H}}(\mbf{Y}|\mbf{X},\mbf{W})$ instead of 
     $\overline{\mbf{H}}(\mbf{Y}|\mbf{X})$. In this case, we have to observe that 
     $$\frac{1}{n}\log\frac{1}{\PR(Y^n|X^n,W^n)}$$
     has a separable form and hence, by ergodicity of $\{W_n\}$ we have that its limit superior
    is $\EX_{\pi}\left[H(N(W))\right]$.
\end{IEEEproof}

We define the queue to be {\em unpredictable} given noise if for all $n$ and $k$
$$\PR(W_{n+k}|N^n)=\PR(W_{n+k})\mbox{,}$$
i.e., $N^n$ provides no information about the future of the queue process.

\begin{theorem}
\label{thm:queueCapacityRC}
For a random bijective queue-channel the capacities are $\lambda\left(\log |\mathcal{X}| - \EX_{\pi}\left[H(N(W))\right]\right)$ and $\lambda\left(\log |\mathcal{X}| - H\left(\EX_{\pi}\left[N(W)\right]\right)\right)$, respectively, when the receiver knows  the arrival and departure times of the symbols and when it does not
(assuming the queue to be {\em unpredictable} given noise in the later case).
\end{theorem}

Next, we state a single letter characterization of the queue-channel capacity for a binary symmetric channel, where the probability of a binary
symbol getting flipped is a function of its waiting time. Formally, for a
function $\phi:[0,\infty) \to [0,0.5]$, $\phi(W_i)$ is the probability that
bit $i$ is flipped. (Unlike erasure channels a bit-flip probability of $0.5$ is the most random/noisy case.) Such a queue-channel is of particular interest, as it arises in our setting when the qubits decohere according to a quantum bit-flip channel, or a depolarising channel~\cite{NCBook}. The following result is a corollary to
Theorem~\ref{thm:queueCapacityRC}.

%Note that the binary symmetric queue-channel is a special case of the reversible channel
%and hence, the capacity result of Thm.~\ref{thm:queueCapacityRC} is applicable
%here. 
\begin{corollary}
\label{thm:queueCapacityBSC}
For a binary symmetric queue-channel the capacity is 
$\lambda\left(1 - \EX_{\pi}\left[H(\phi(W))\right]\right)$
or $\lambda\left(1 - H\left(\EX_{\pi}\left[\phi(W)\right]\right)\right)$,
depending on whether the receiver knows arrival and departure times or not {(assuming
the queue to be {\em unpredictable} given noise in the later case)}.
\end{corollary}

As discussed in Sec.~\ref{sec:EC} an exponential functional relationship between error and waiting time is relevant to the 
decohering quantum channels and the age of information in packet communication. Hence, it is important to understand the dependence of the queue-channel 
capacity on service statistics in this case.

\begin{theorem}
\label{thm:bestServiceBSCnoCSIR}
For binary symmetric queue-channel  with
$\phi(W)=\frac{1}{2}\left(1-\exp(-\kappa W)\right)$ where the receiver
does not know the arrival and departure times and {the queue is {\em unpredictable} given noise}, at any $\lambda$ the deterministic service time maximizes the capacity over all
service time distributions.
\end{theorem}

Therefore, service time jitter is undesirable in this channel too.

\section{Concluding Remarks and Future Work}

In this paper, we used simple queue-channel models to  characterize the capacity of channels with waiting time dependent errors. Though our main motivation stems from quantum communications, where we characterize the rate at which classical information can be transmitted using orthogonal quantum states that decohere in time, the model closely captures scenarios in crowdsourcing and multimedia streaming.  

We believe there is ample scope for further work along several directions. Firstly, it is important to move away from the restriction of  using only orthogonal states at the encoder and a fixed measurement at the receiver, and allow for arbitrary superposition states at the encoder, and arbitrary measurements at the receiver. This would allow us to invoke the true classical capacity of the underlying (non-stationary) quantum channel, in terms of a quantity~\cite{HN03_genCapacity} analogous to the classical inf-information rate. It remains an interesting technical challenge to obtain a formula for the queue-channel capacity in this general scenario, and identify channels for which the classical coding strategy would still be optimal. Furthermore, we can also consider other widely studied quantum channel models, such as the phase damping and amplitude damping channels. 

We have only considered uncoded quantum bits in this paper. We can also quantitatively evaluate the impact of using quantum codes to protect qubits from errors. Employing a code would enhance robustness to errors, but would also increase the waiting time due to the increased number of qubits to be processed. It would be interesting to characterise this tradeoff, and identify the regimes where using coded qubits would be beneficial or otherwise.

More broadly, we believe our work highlights the importance of explicitly modelling delay induced errors in quantum communications. As quantum computing takes strides towards becoming an ubiquitous reality, we believe it is imperative to develop processor architectures and algorithms that are informed by more quantitative studies of the impact of delay induced errors on quantum information processing systems.

\bibliographystyle{IEEEtran}
\bibliography{conf_abrv,abrv,lrv_lib}

%\appendix
%\subsection{Proof of Lemma \ref{lem:MG1opti}}
%
%We show that the capacity is a concave function of $\lambda$ over $[0,1)$.
%Note that to show a function is concave it is sufficient to show that the function plus a linear
%function is concave.
%
%Then, let us consider the function 
%$$ \frac{\lambda (1-\lambda)}{1 - \alpha \lambda} - \frac{\lambda}{\alpha},$$
%which becomes $\frac{\left(1-\frac{1}{\alpha}\right)\lambda}{1-\alpha\lambda}$. 
%
%Note that for any
%$0<a<1$ the function $\frac{x}{1-ax}$ is convex in $x$. This can be seen by adding $\frac{x}{a}$ to it and observing that this gives $\frac{1}{1-ax}$ which is convex in $x$ for $0\le x < \frac{1}{a}$. As $\alpha<1$ this implies that $\frac{\left(1-\frac{1}{\alpha}\right)\lambda}{1-\alpha\lambda}$ is concave in $\lambda$.

\end{document}